\documentclass[conference]{IEEEtran}
\IEEEoverridecommandlockouts
\PassOptionsToPackage{table}{xcolor}

\usepackage{cite}
\usepackage{amsmath,amssymb,amsfonts}
\usepackage{algorithm}
\usepackage{algpseudocode}
\usepackage{graphicx}
\usepackage{textcomp}
\usepackage{xcolor}

\usepackage{tabularx}
\usepackage{booktabs}
\usepackage{xcolor}
\usepackage{caption}

\PassOptionsToPackage{hyphens}{url}

\def\BibTeX{{\rm B\kern-.05em{\sc i\kern-.025em b}\kern-.08em
    T\kern-.1667em\lower.7ex\hbox{E}\kern-.125emX}}

\makeatletter
\def\ps@IEEEtitlepagestyle{%
  \def\@oddfoot{\mycopyrightnotice}%
}
\def\mycopyrightnotice{%
  \begin{minipage}{\textwidth}
  \centering \scriptsize
  Copyright~\copyright~2024 IEEE. Personal use of this material is permitted.  Permission from IEEE must be obtained for all other uses, in any current or future media, including reprinting/republishing this material for advertising or promotional purposes, creating new collective works, for resale or redistribution to servers or lists, or reuse of any copyrighted component of this work in other works.
  \end{minipage}
}
\makeatother

\begin{document}

\title{VELLET: Verifiable Embedded Wallet for Securing Authenticity and Integrity\\
}

\author{\IEEEauthorblockN{Hiroki Watanabe}
\IEEEauthorblockA{\textit{The Japan Research Institute, Ltd.} \\
Tokyo, Japan \\
watanabe.hiroki@jri.co.jp}
\and
\IEEEauthorblockN{Kohei Ichihara}
\IEEEauthorblockA{\textit{The Japan Research Institute, Ltd.} \\
Tokyo, Japan \\
ichihara.kohei@jri.co.jp}
\and
\IEEEauthorblockN{Takumi Aita}
\IEEEauthorblockA{\textit{The Japan Research Institute, Ltd.} \\
Tokyo, Japan \\
aita.takumi.m2@jri.co.jp}
}

\maketitle

\begin{abstract}
The blockchain ecosystem, particularly with the rise of Web3 and Non-Fungible Tokens (NFTs), has experienced a significant increase in users and applications. However, this expansion is challenged by the need to connect early adopters with a wider user base. A notable difficulty in this process is the complex interfaces of blockchain wallets, which can be daunting for those familiar with traditional payment methods. To address this issue, the category of "embedded wallets" has emerged as a promising solution. These wallets are seamlessly integrated into the front-end of decentralized applications (Dapps), simplifying the onboarding process for users and making access more widely available. However, our insights indicate that this simplification introduces a trade-off between ease of use and security. Embedded wallets lack transparency and auditability, leading to obscured transactions by the front end and a pronounced risk of fraud and phishing attacks. This paper proposes a new protocol to enhance the security of embedded wallets. Our VELLET protocol introduces a wallet verifier that can match the audit trail of embedded wallets on smart contracts, incorporating a process to verify authenticity and integrity. In the implementation architecture of the VELLET protocol, we suggest using the Text Record feature of the Ethereum Name Service (ENS), known as a decentralized domain name service, to serve as a repository for managing the audit trails of smart contracts. This approach has been demonstrated to reduce the necessity for new smart contract development and operational costs, proving cost-effective through a proof-of-concept. This protocol is a vital step in reducing security risks associated with embedded wallets, ensuring their convenience does not undermine user security and trust.
\end{abstract}
\begin{IEEEkeywords}
blockchain, embedded wallet, decentralized applications, security audit
\end{IEEEkeywords}
\section{Introduction}
The blockchain-based ecosystem continues to attract a vast array of new participants, following the rise of Web3 and the Non-Fungible Token (NFT) market. Current estimates suggest that over 420 million users possess cryptographic assets \cite{TripleA2023}. As of December 2023, it is observed that there are approximately 15,000 decentralized applications (Dapps) and in excess of 431,000 smart contracts identified across upwards of 62 blockchains \cite{DappRadar2023}. However, in order to achieve further growth, it is necessary to overcome the gap between the early adopters and the early majority, known as the "chasm".

Crossing this chasm faces several technical challenges. A prominent barrier is the complex user interfaces related to blockchain-specific services, such as wallets. For instance, Metamask\cite{metamask2023}, a popular non-custodial wallet among cryptocurrency enthusiasts, can be challenging for users accustomed to conventional payment systems. Recent extensive user interaction studies have revealed difficulties with wallets, including specific issues in the crypto-currency domain, such as transaction complexities, fees, address, and key management \cite{Froehlich2022}. Many of these problems could be addressed by emulating the design of existing online banking and payment systems, which users are already familiar with \cite{Voskobojnikov2021}.

To address these challenges, the emerging category of embedded wallets has attracted significant attention\cite{zengo2023personalvsembedded}, with multiple companies now entering the distribution phase of these products\cite{thirdweb2023embeddedwallet, privy2023embeddedwallets, dynamic2023embeddedwallets}. These are seamless, non-custodial wallets instantly generated for blockchain applications. Unlike traditional non-custodial wallets, which exist independently outside a specific app but can connect to it, embedded wallets create a user's wallet in the background using conventional login credentials (such as email or social logins) when creating an account in the app. This eliminates the need to first create a wallet and then connect it to a Dapp account, thus favoring user onboarding to Dapp services.

However, there is generally a trade-off between usability and security. Our insights suggest that the simple and user-friendly appearance of embedded wallets could induce scam and phishing in wallet services. The most significant flaw is that embedded wallets do not offer transparency and auditability, making it impossible to prevent fraudulent activities. Users have no choice but to trust the representation of the embedded wallet displayed on the web application. Cryptocurrencies or NFTs they believe they have purchased may only appear on the embedded wallet and do not actually exist on the blockchain. Additionally, users may not immediately realize if the purchase amount has been manipulated to differ significantly from the order book.

\begin{figure*}[tbp]
\centering
\includegraphics[width=\textwidth]{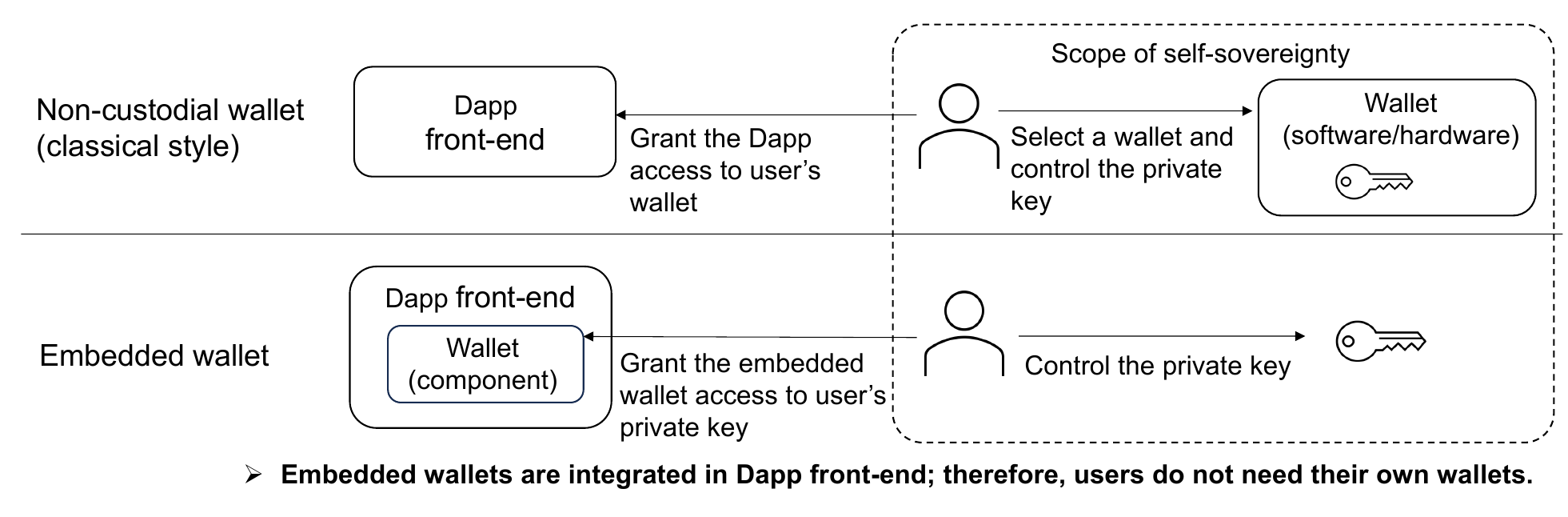}
\captionsetup{width=\textwidth,justification=justified}
\caption{Comparison of Embedded Wallets and Conventional Non-Custodial Wallets}
\label{fig:fig1}
\end{figure*}
In this paper, we propose a new wallet verification protocol that incorporates verifiability into embedded wallets. The protocol ensures that the wallet has been audited by an audit organization, such as a security audit company or a community of security experts and enthusiasts. Additionally, by linking the integrity of the embedded wallet with the domain of the Dapp front-end provided, it prevents tampering with the embedded wallet by external attackers. A distinctive implementation idea is to use the Text Records\cite{ENSIP5} of the Ethereum Name Service (ENS)\cite{ENSWebsite}, a decentralized name service for Ethereum, as a repository that links the domain to the integrity of the resources of the wallet. This method is cost-effective as it ensures transparency while reducing the expenses associated with introducing and operating new contracts.

\section{Background}

\subsection{Non-Custodial and Embedded Wallets}

According to the taxonomy of wallets proposed by Erinle et al.\cite{erinle2023sok}, wallets are primarily categorized based on two factors: "control over assets" and "internet connectivity". The term "control over assets" refers to whether the wallet's private key is custodial, stored by a third party, or non-custodial, directly controlled by the user. Conversely, "internet connectivity" distinguishes between hot wallets, which are always connected to the internet, and cold wallets, which are operated on physical hardware isolated from the internet.

Given the importance of sovereignty in crypto assets, non-custodial wallets are often preferred by users. These wallets are accessible through various means, including browser extensions or standalone applications. Notable examples include MetaMask\cite{metamask2023}, Argent\cite{argent2023}, and Trust Wallet\cite{trustwallet2023}. From the service provider's perspective, there is significant interest in hot wallets, as they facilitate user onboarding to their services without barriers. 

Consequently, the emergence of cloud-based Wallet as a Service (WaaS) provided by cryptocurrency companies such as Coinbase\cite{coinbase2023waas} and Circle\cite{circle2023programmablewallets} is noteworthy. WaaS is a cloud-based service that delivers essential wallet functionalities, designed for seamless integration into the applications and services of businesses and developers. It provides APIs that enable the incorporation of wallet capabilities into business applications, primarily in non-custodial and hot wallet configurations. In terms of key management, WaaS frequently employs technologies such as Multi-Party Computation (MPC) to ensure that private keys remain under the control of the user, contrasting sharply with custodial solutions where key management is handled by third-party entities.

Embedded wallets constitute a specific category of WaaS. Targeted primarily at web application developers, these wallets are available as integrated components, including a JavaScript SDK and a user interface. This configuration allows developers to effortlessly embed them into their web service front-end, thus facilitating easy access to blockchain services for users. Figure \ref{fig:fig1} illustrates a rudimentary comparison between the design of a conventional non-custodial wallet and an embedded wallet. Embedded wallets contribute to improved user onboarding by simplifying the use of Dapps, reducing the extent of user control from the entire wallet to just the private key. Notable companies active in this field include Thirdweb\cite{thirdweb2023embeddedwallet}, Privy\cite{privy2023embeddedwallets}, and Dynamic\cite{dynamic2023embeddedwallets}. Since there is no clear definition of embedded wallets in the existing literature, we refer to their products and define them as follows in this paper:

\begin{itemize}
  \item Offered as a non-custodial, internet-connected web wallet.
  \item A single wallet is provided for a single web application (different wallet spaces are provided for different web apps).
  \item Can be easily integrated as part of one's own app in the form of reusable components.
  \item The web app and wallet are tightly coupled, and exporting the private key is necessary to transfer it to an external wallet.
  \item The implementation method for key storage is not limited; various methods such as MPC, key sharding, and contract wallets are seen in practice. Some services allow the developers of the web application themselves to choose.
\end{itemize}
\begin{figure*}[tbp]
\centering
\includegraphics[width=0.8\textwidth]{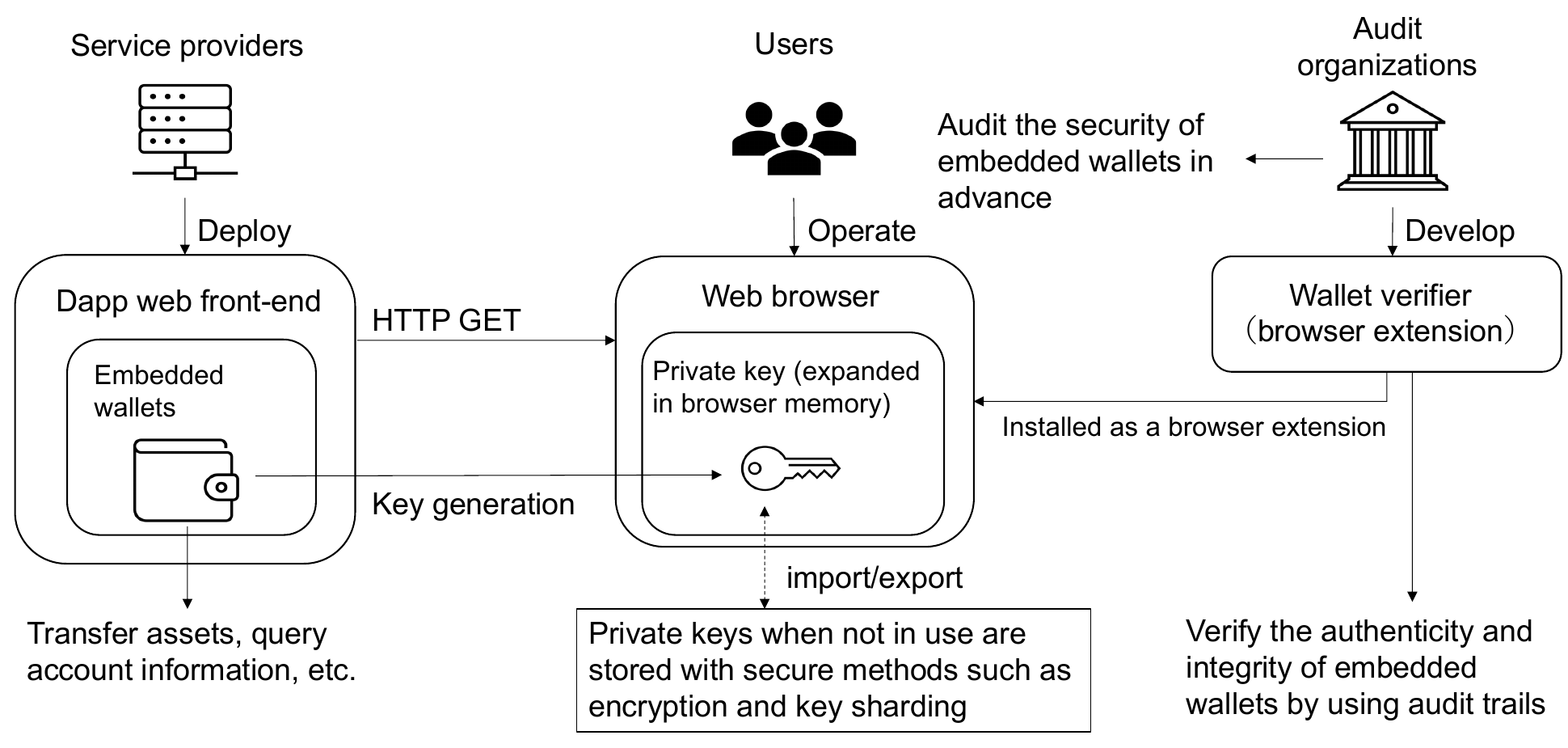}
\captionsetup{width=0.8\textwidth,justification=justified}
\caption{Interactions between Service Providers, Users, and Audit Organizations}
\label{fig:fig2}
\end{figure*}
\subsection{Challenges of Embedded Wallets}

In 2022, the cryptocurrency trading sector experienced substantial losses due to fraudulent activities, which amount to approximately \$5.9 billion\cite{Chainalysis2023}.
A significant threat within this domain is phishing, wherein cybercriminals use deceptive emails or websites to illicitly acquire users' cryptocurrencies\cite{andryukhin2019phishing} or NFTs\cite{yang2023trail}. Although phishing strategies range from social engineering tactics like clone attacks to technical approaches such as DNS hijacking\cite{andryukhin2019phishing}, these methods typically involve mimicking legitimate transactions to deceive users.

A proactive measure against such phishing in crypto assets involves meticulous verification of wallet displays\cite{trustwallet2023phishing}. These displays provide essential information, including transaction details, recipient addresses, and the amount being transferred. Vigilant examination of this information can alert users to potential irregularities. Additionally, some wallets are equipped with features to identify and warn against fraudulent accounts, enhancing security\cite{metamask2023blockaid,trustwallet2023securityscanner}. This is particularly effective as conventional non-custodial wallets, which operate independently from Dapp front-ends, remain under user control.

Conversely, embedded wallets, which provide user interfaces akin to those in Web 2.0, can potentially enable more sophisticated phishing attacks. Compromised or counterfeit embedded wallets can alter or misrepresent crucial transaction information, thus mimicking the appearance of legitimate wallets. They might also falsely claim to have passed security audits to gain user trust. The core issues can be summarized as follows:

\begin{itemize}
  \item Difficulty in distinguishing between legitimate and malicious embedded wallets.
  \item Inability of users to detect when an embedded wallet is compromised due to tampering with a legitimate Dapp's front-end.
\end{itemize}

These issues highlight challenges in ensuring both authenticity and integrity, which have long been central in digital security. To effectively address these challenges, various mechanisms and technologies have been developed historically. For instance, DNSSEC (Domain Name System Security Extensions)\cite{rfc2535} offers a suite of extensions to DNS, providing clients with origin authentication, authenticated denial of existence, and data integrity. It employs public-key cryptography to confirm that DNS records remain unaltered, thereby preserving their authenticity and integrity. In DNSSEC, the root zone's public key is a common trust anchor, validating the authenticity of the entire root zone. Similarly, Code Signing\cite{cooper2018security} allows software distributors to certify the authenticity and integrity of their code. By digitally signing executables and scripts, developers can affirm that their code has not been modified or compromised. When combined with security audits, such as malware scans, it's feasible to distribute applications verified for trustworthiness and compatibility (e.g., Windows Driver Signing \cite{microsoft_driver_signing}, macOS Code Siging \cite{apple_code_signing}, Android App Signing \cite{android_app_signing}). Here, the trust anchor is typically the platform operator managing the application distribution infrastructure, with client devices relying on the operator to verify code signatures.

Historically, operations ensuring authenticity and integrity have involved a third-party trust point, in addition to the data provider and recipient. This party validates the provider's data through mechanisms like digital signatures. However, in emerging digital frameworks employing blockchain technology, such as Web3, the move towards decentralization often leads to a dearth of mechanisms for ensuring authenticity and integrity. Embedded wallets in this context exemplify this issue, which remains largely unaddressed in existing literature.

\section{VELLET Overview}

\subsection{Objective}

This paper proposes the VELLET protocol, aimed at addressing the issues of authenticity and integrity within embedded wallets. VELLET seeks to fulfill the following requirements:

\begin{itemize}
  \item Users can receive security warnings when using embedded wallets that have not been audited, addressing authenticity.
  \item Users can receive security warnings if the embedded wallet they are about to use has been compromised or tampered with, addressing integrity.
  \item The system is decentralized, transparent, and can be easily implemented.
\end{itemize}

\subsection{Entities}
The implementation of the VELLET protocol is realized through the cooperation of three principal entities. Figure \ref{fig:fig2} illustrates the relationship between the entities utilizing the VELLET protocol and the software they control. The assumptions in this paper are as follows:

\begin{enumerate}
\item \textbf{Service Providers.}
  Service providers provide decentralized applications (Dapps) linked to the blockchain. Their front-end employs embedded wallets, providing users with a means to connect to the blockchain. While embedded wallets may be offered via Wallet as a Service (WaaS), the sovereignty of the system rests with the service providers. Therefore, they hold the rights to control both the Dapp front-end and the embedded wallet.
  
\item \textbf{Users.}
  Users access the service provider's Dapp front-end through a Web browser. They manage the private keys used within the embedded wallet. It is important to note that while service providers supply the embedded wallet, they do not possess the private keys; hence, they are non-custodial. Users import their keys onto the embedded wallet and deploy them in the browser's memory only when using the wallet. Thus, only the user possesses and controls the rights to their private keys.
  
\item \textbf{Audit Organizations(e.g., security audit companies).}
 Audit organizations conduct security audits on the trustworthiness of embedded wallets and grant their approval. In the blockchain service industry, it's common for numerous companies to perform third-party audits to assess the security and soundness of smart contract code. However, the records or documentation of such audit trails are not widely available for public scrutiny. In VELLET, audit organizations submit an audit trail to the blockchain beforehand, and this trail is used to carry out the verification process for embedded wallets. VELLET introduces a new software tool called Wallet Verifier. Installed as a browser extension for the user, the Wallet Verifier utilizes audit trails to confirm the authenticity and integrity of embedded wallets. This paper posits that audit organizations are responsible for developing and maintaining the Wallet Verifier, thus preventing the centralization of authority with the service providers. An example of an auditing organization could be a blockchain-specialized security audit company\cite{certik2023, hacken2023}, or a community of security experts.
\end{enumerate}

\subsection{Architecture}
\begin{figure}[tbp]
\centering
\includegraphics[width=\columnwidth]{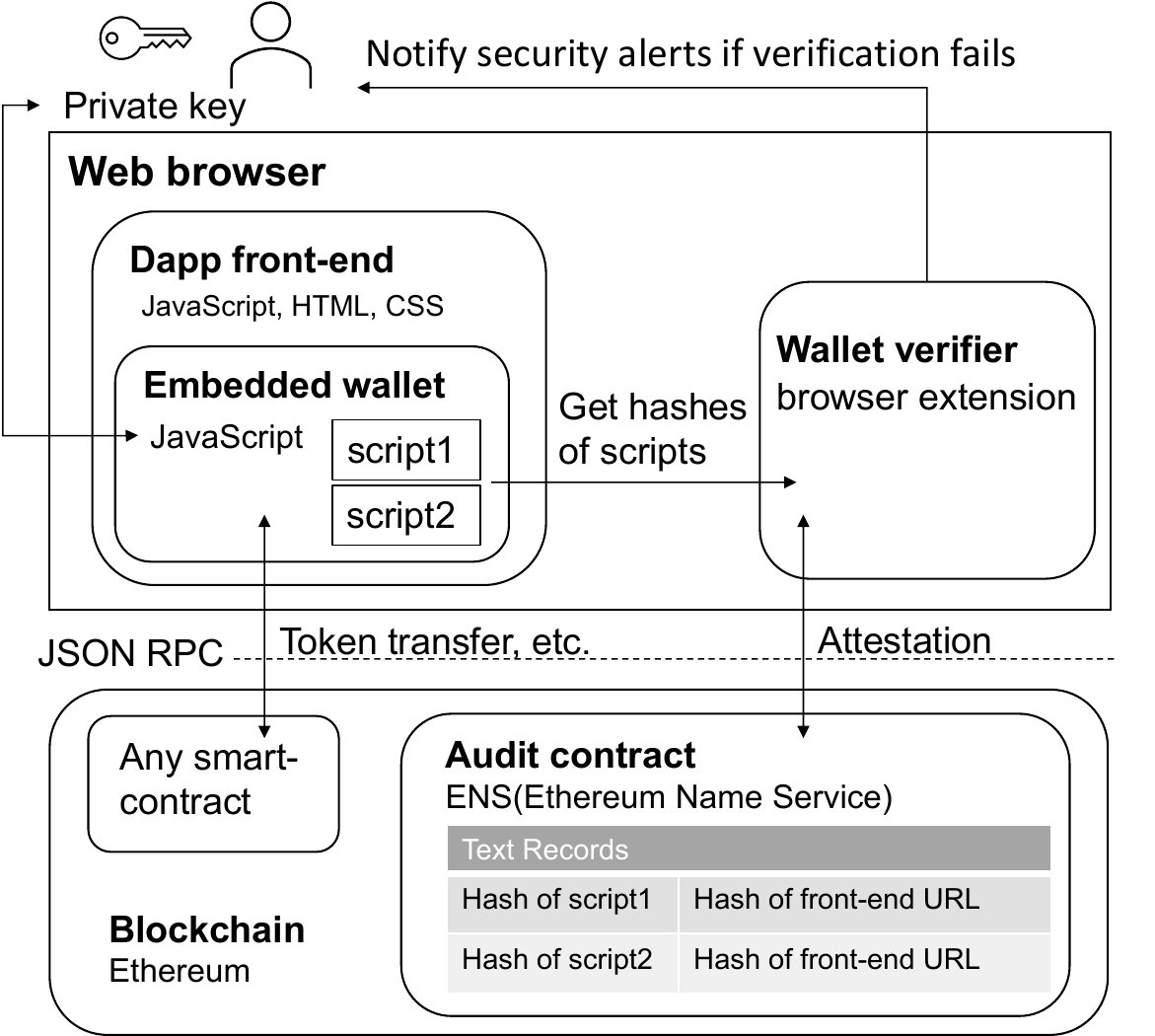}
\caption{Overview of the software architecture executing the proposed protocol.}
\label{fig:fig3}
\end{figure}

We describe the overall architecture of the software modules necessary for the VELLET verification process. An overview of this architecture is shown in Figure \ref{fig:fig3}.

\textbf{Embedded Wallet.}
The embedded wallet software $w$ is integrated within the Dapp front-end $D$, meaning $w \in D$. This web-based wallet $w$ is optimally adjusted for the use of Dapp $D$ and may have restricted functions compared to common non-custodial wallets such as Metamask. For instance, it might only generate private keys or receive tokens, omitting the functionality to sign payment transactions. (Specific use cases for this restricted embedded wallet are explained in detail in Section V-A.) $w$ consists of JavaScript modules, and during the verification process, some fragmented scripts within this module are subject to inspection.

\textbf{Wallet Verifier.}
While embedded wallets help in user onboarding and provide excellent user experience, there is a risk that it could evolve into a centralized structure. The service provider that offers Dapp and hosts the embedded wallet could conceal most of the internal operations. Additionally, it is vulnerable to recognized attacks such as website changes due to Dapp hacking or phishing attempts that redirect users to fake sites. To counter these drawbacks, we introduce a verification process for the embedded wallet. Each time a user accesses the embedded wallet, the Wallet Verifier scrutinizes the wallet code and certifies its functionality. In particular, the embedded wallet is only executed if the code has not been tampered with and if code execution conditions (such as in encrypted communication) are met. This wallet verifier is designed as a browser extension that can analyze JavaScript modules on the Dapp front-end and the embedded wallet.
Furthermore, the wallet verifier communicates with the audit contract, a smart contract on the blockchain, to attest to the integrity and authenticity of the embedded wallet. Attestation is performed by matching the script's hash against the audit trail previously submitted to the audit contract by the audit organization. (Details are discussed in Section IV.)

\textbf{Audit Contract with ENS Text Records.}
Audit contract $C$ is a type of smart contract executed on the blockchain. Its main purpose is to manage the audit trails of embedded wallets and to ensure their authenticity and integrity. Ideally, the registration of audit trails in the contract should be conducted by a third-party audit organization, following the principles of decentralization, excluding the service provider. The audit trail is composed of essential audit-related data. This includes the URL (Uniform Resource Locator) of the Dapp front-end, which hosts the embedded wallet, and the hashes of the scripts used within the wallet. The audit contract publishes the audit trails on the blockchain for reference by other participants. This enables users and blockchain-connected wallet verifiers to inspect the audit results of the embedded wallet and evaluate its trustworthiness.

The point to be emphasized in our approach is the proposal to adopt the Ethereum Name Service (ENS), a decentralized naming service running on the Ethereum blockchain, as the foundational structure of the audit contract, instead of constructing and operating an audit contract from scratch. ENS, which consists of smart contracts, is a decentralized domain name system operational on the Ethereum blockchain, capable of holding text records of specific metadata and general information, not just domain names\cite{ENSIP5}. Registering information with ENS operating on the blockchain requires a signed transaction, ensuring the authenticity of the registrar as the domain owner. In this manner, an auditing organization can fulfill the same requirements as constructing an audit contract by registering audit trails as text records linked to domains. Utilizing ENS allows auditing organizations to reduce the costs of independently operating an audit contract and lowers the threshold for adopting the VELLET protocol.

\section{Protocol in detail}
In this section, we explain the detailed workings of the VELLET protocol, which operates on the aforementioned architecture. The VELLET protocol is made up of two fundamental phases:

\subsection{Audit Phase}
For the sake of simplicity, this document describes the audit contract $C$ using the most basic implementation. $C$ consists of a key-value data store $C.\textit{store}$ for storing audit trails, a stateless operation $C.\textit{regit()}$ for registering audit trails $A_t$ with $C.\textit{store}$, and an operation $C.\textit{getAudit()}$ for referencing $A_t$. The audit contract $C$ is deployed on any blockchain that supports the execution of smart contracts. Only an audit organization with appropriate permissions can perform write operations that involve state changes to the audit contract $C$. The audit organization reviews the program code of the embedded wallet $w$ and inspects its behavior during the execution of the code. A code audit contains the selection of appropriate cryptographic primitives, the use of secure libraries, and the absence of inappropriate external connections. The audit trail set, denoted as $A_t$, is assured to encompass at least the hash $H_u$ of a Uniform Resource Locator (URL) string $u$, which identifies the location of the Dapp front-end; in other words, $H_u \in A_t$. Both the hash value $H_u$ and $H_w$, representing the hash value of the embedded wallet $w$'s code, are stored in the contract $C$ as key-value pairs. Thus, the data store $C.\textit{store}$ is updated as follows:
\begin{equation}
{C}.\textit{store} \leftarrow \textit{TX}_b(C.\textit{regit}(H_w, A_t))
\end{equation}
Here, the execution of $C.\textit{regit}(H_w, A_t)$ is performed through a blockchain transaction. That is, a valid execution transaction $\textit{TX}_b(C.\textit{regit}(H_w, A_t))$ on the blockchain network is broadcasted by an audit organization with registration permissions for the contract $C$. Through the execution process of the smart contract, the transaction is processed, and the pair $(H_w, A_t)$ is stored in the data store $C.\textit{store}$ defined in contract $C$. Using $H_w$ as a query, the audit trails $A_t$ can be retrieved as follows:
\begin{equation}
{C}.\textit{getAudit}(H_w) \rightarrow A_t \quad \text{where} \quad H_u \in A_t
\end{equation}
The $\textit{getAudit()}$ operation retrieves the audit trail set $A_t$ from the store without changing the state of the smart contract. It does not require issuing a transaction to read the audit trails, and the value is publicly accessible. It should be noted that $A_t$ contains the hash $H_u$ of the URL $u$ indicating the location of the Dapp front-end. The preimage of the pair $(H_w, H_u)$ registered with $C$ is $(w, u)$, which serves as the trust anchor between the embedded wallet and the location where the Dapp front-end is hosted. Furthermore, $A_t$ can include additional audit-related information, such as the audit completion date or the name of an audit company, but this document does not provide specific definitions for these to ensure clarity.
\subsection{Verification Phase}

The previous subsection explained how the trust anchor of the embedded wallet $w$ is established within the audit contract $C$ during the audit phase. In this section, we propose a protocol that utilizes this trust anchor to enable users to safely use the wallet $w$. As mentioned in Section III-C, the user's client terminal requires software for the wallet verification. A wallet verifier $\textit{Ins}$ is delivered to the client from a repository by an entity different from the Dapp service provider and is installed on the client. $\textit{Ins}$ incorporates the operations $\textit{verify()}$ for wallet verification and $\textit{execute()}$ for code execution. The $\textit{verify()}$ operation returns the result of wallet verification as follows:
\begin{equation}
\textit{Ins.verify}(w, u, \textit{ca}, \textit{cond}) \in \{ \text{true}, \text{false} \}
\end{equation}
Here, $ca$ indicates the access information for the audit contract, such as the contract's address or Application Binary Interface (ABI). $\textit{cond}$ represents additional verification conditions. For example, it can be used to verify the name of the audit agency or the execution environment. The verification uses the audit contract as a trust anchor to verify that the wallet $w$ and the URL $u$ indicating the provider have been audited. Furthermore, this verification process ensures that the wallet $w$ has not been tampered with, demonstrating resistance to phishing attacks. Additionally, the execution permission for wallet $w$ resides in the wallet verifier $\textit{Ins}$. In other words, $\textit{Ins}$ implements the following function:
\begin{equation}
\textit{Ins.execute}(w) \rightarrow \text{result}
\end{equation}

The $\textit{execute()}$ function executes the functionalities provided by the embedded wallet $w$. Importantly, the execution permission is granted to $\textit{Ins}$, not to the Dapp front-end where the wallet is embedded. By decentralizing permissions among entities, this approach eliminates a single point of failure and distributes security risks, enhancing overall security of the system. In contrast to conventional feature-rich wallets, wallets in this context are designed with simplicity in mind and only possess the necessary functionalities. Furthermore, by applying the principles of modular design, components such as key generation can be provided as reusable separate modules. Therefore, $w$ can be embedded in the Dapp as several modules represented by ${w_i, i=0,1,2...}$. In such cases, Equation (3) can be extended as follows:

\begin{equation}
\text{Verification} =
\begin{cases}
\text{true} & \text{if } \forall i, \textit{Ins.verify}(w_i, u, \textit{ca}, \textit{cond}) = \text{true} \\
\text{false} & \text{otherwise}
\end{cases}
\end{equation}
In our implementation, discussed in Section V, the wallet verifier $\textit{Ins}$ alerts the user with a security warning when the verification result is false. This mechanism provides resistance against fraudulent Dapp websites by informing users of potential risks such as wallet tampering $w_i$ or inadequate URL auditing $u$. 
%

\section{Implementation}

To verify the feasibility of the proposed protocol, we worked on its implementation. In this section, we explain the details of the implementation to facilitate a comprehensive understanding of the protocol.

\subsection{Applications and Scenarios}

Embedded wallets are designed to make using Dapps as easy as traditional Web 2.0 apps, aiming to boost their widespread adoption. A key application in this context is the distribution of utility-focused NFTs (Non-Fungible Tokens), moving away from speculative trading. In our VELLET proof-of-concept, we explore using utility-based NFTs for digital identity. As described by Weyl et al.\cite{Weyl2022}, Soulbound Tokens (SBTs) are non-transferable NFTs employed to articulate the personal attributes such as identity, skills, experiences, and qualifications of the individual to whom they are assigned. This allows for the potential use of SBTs as a means to bridge digital identities, social bonds, and trust verification between the blockchain economy and the societal context. Accordingly, we have constructed a scenario that leverages embedded wallets for the deployment of SBTs via a Dapp.

\textbf{Use Case:}
The user is a member of a specific community (e.g., an art club, sports club, or an open-source developer community). The community site issues SBTs to its members as proof of membership. The user can present SBT to third-party institutions outside the community site. Unlike credit scoring systems provided by centralized entities, SBTs demonstrates bottom-up social creditworthiness. Third-party institutions can evaluate the user's activities within the community and provide certain benefits (such as unsecured loans, voting rights, job referrals, etc.) based on trust relationships.

\textbf{Scenario I: Distribution of SBTs}
\begin{enumerate}
\item A user logs in to the website of a community.
\item After logging in, the user requests the issuance of a membership SBT from their profile page.
\item The site generates an encrypted private key through the embedded wallet.
\item The SBT is issued to the address associated with the private key.
\item The user exports the private key.
\end{enumerate}

\textbf{Scenario II: Presentation of SBTs}
\begin{enumerate}
\item The user accesses a website dedicated to presenting or submitting SBTs.
\item The user imports the encrypted private key into the embedded wallet of the website.
\item The embedded wallet generates an electronic signature and sends it to the website.
\item The website verifies and authenticates the signature.
\end{enumerate}

\subsection{Implementation of Proof-of-Concept}
We have developed an implementation for a proof-of-concept of VELLET based on the scenario we devised. For the web front-end of the Dapp, we have chosen the Next.js framework, and for the issureing of SBTs, we are utilizing Ethereum's Goerli testnet. The SBT standard adopts ERC-5192\cite{EIP5192} (Minimal Soulbound NFTs) and is backward compatible with the well-known NFT standard, ERC-721\cite{EIP721}. We have adopted ethers.js for the construction of the embedded wallet. Ethers.js is a JavaScript library that allows developers to interact with the Ethereum blockchain and its ecosystem, providing wallet functionality (such as private key generation and transaction signing) that operates with JavaScript on a browser. The wallet verifier has been developed to adhere to the standards of Chrome Extension Manifest V3. The audit contract with which the wallet verifier interacts, as mentioned in Section III, utilizes the ENS (Ethereum Name Service) Text Records\cite{ENSIP5}. In this implementation, we used the ENS deployed on the Goerli testnet.

Let us clarify the parts related to the VELLET protocol among the scenarios. As mentioned above, the VELLET protocol is realized through the collaboration of the embedded wallet, audit contract, and wallet verifier. The embedded wallet is used in Scenarios I-3 to I-5 and Scenarios II-2 to II-3 steps. In these scenarios, the embedded wallet operates only when cryptographic operations involving the user's private key are required. The wallet is seamlessly integrated into the application, naturally embedded within the user interface interaction flow of the application. In the specific implementation, the embedded wallet is described as a client-side JavaScript module executable within the browser and called within the Dapp's website. The critical aspect to emphasize is the cryptographic operations are exclusively executed within the user's browser, never on the server side, ensuring that the embedded wallet is inherently non-custodial.

The audit contract, which employs ENS's Text Records \cite{ENSIP5}, serves as a searchable data store. It facilitates the retrieval of audit trails for embedded wallets by using the hash value of the Dapp front-end site URL as a query. The audit trails are audit results in which an audit company has pre-inspected the module code of each embedded wallet, as described in the audit phase in Section IV-A. Examples of audit points at this stage include the following:
\begin{itemize}
\item Selection of appropriate cryptographic primitives.
\item Setting appropriate seed values for key generation.
\item Prevention of content being sent to external servers (e.g., not sending the private key).
\end{itemize}

\begin{algorithm}[tbp]
\caption{Script Verification in Wallet Verifier}
\begin{algorithmic}[1]
\Procedure{Extract and Verify Scripts}{}
    \State $\textit{tabId} \gets \textit{getActiveTabId()}$
    \State $\textit{tabUrl} \gets \textit{getActiveTabUrl(tabId)}$
    \State $\textit{pageUrlHash} \gets \textit{getHash(tabUrl)}$
    \State $\textit{html} \gets \textit{getInnerHTMLFromTab(tabId)}$
    \State $\textit{pageDoc} \gets \textit{parseHTML(html)}$
    \State Extract scripts from $\textit{pageDoc}$ and save in $\textit{scripts}$
    \For{each $\textit{script}$ in $\textit{scripts}$}
       \State Compute hash of $\textit{script}$ and append to $\textit{scriptDigests}$
    \EndFor
    \State $\textit{provider} \gets \textit{connectToWeb3Provider()}$ \\
    \Comment The following example uses ENS as an audit contract.
    \State $\textit{ens} \gets \textit{configureENS(provider)}$
    \State $\textit{domain} \gets 'vellet.eth'$
    \For{each $\textit{digest}$ in $\textit{scriptDigests}$}
        \State Get text record for $\textit{digest}$ from ENS and save result in $\textit{scriptChecks}$
        \State $\textit{textRecord} \gets \textit{ens.name(domain).getText(digest)}$
        \State $\textit{isValid} \gets (\textit{textRecord} == \textit{pageUrlHash})$
        \State Append $\textit{isValid}$ to $\textit{scriptChecks}$
    \EndFor
    \If{$\textit{scriptChecks}$ contains any $\textit{false}$ value}
        \State $\textit{setIsNotSecure(true)}$
    \Else
        \State $\textit{setIsNotSecure(false)}$
    \EndIf
\EndProcedure
\end{algorithmic}
\end{algorithm}
The main function of the wallet verifier, implemented as a browser extension, is to attest to the legitimacy of the embedded wallet and then initiate wallet operations. This process of attestation is crucial to prevent execution in potentially fraudulent embedded wallets or on counterfeit Dapp websites. Although several implementation approaches are possible, in our design, the wallet verifier, when deployed as a browser extension, processes the HTML document to extract the wallet code encapsulated within script tags. This code, as previously indicated, is modularized as a JavaScript component and is processed as a string in our algorithm. Algorithm 1 presents the pseudocode for this wallet verifier. Given its design as a browser extension, the verifier has the capability to read the HTML Document of the currently active browser tab. It fetches the audit outcomes for the vetted Dapp by using the hash of the embedded wallet code within the HTML as the reference key. The audit data incorporates the hash of the Dapp's URL, facilitating the cross-checking of the Dapp website URL being accessed by the user against its audited status.

\begin{figure}[tbp]
\centering
\includegraphics[width=\columnwidth]{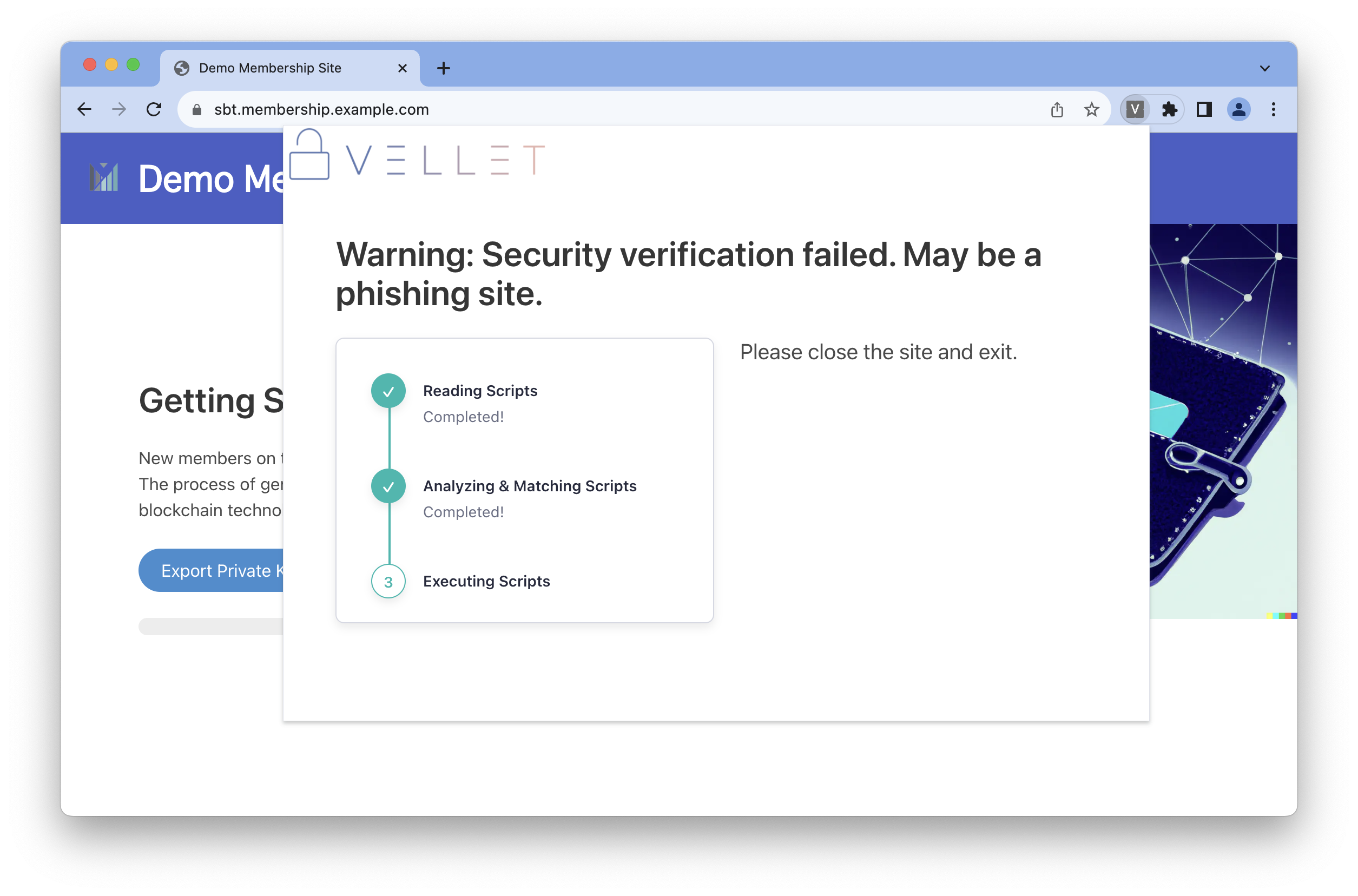}
\caption{The user interface of the wallet verifier, which pops up on the website, displays a warning when the verification of an embedded wallet fails.}
\vspace*{-0.4cm}
\label{fig:fig4}
\end{figure}

If the audit trails of Dapps cannot be obtained, that is, if the hash value of the Dapp URL is not registered on the ENS Text Record, this means the authenticity of the embedded wallet is not secured. Furthermore, if the audit trails are obtained and they do not include the code hash of the embedded wallet, it indicates a loss of the wallet's integrity. Therefore, ENS records are necessary to act as a trust anchor. The resistance to data tampering and the authenticity provided by blockchain infrastructure further strengthen this assertion. In our proof-of-concept, as mentioned in Section III, if either authenticity or integrity is negated, a warning is displayed to the user as shown in Figure \ref{fig:fig4}.

\subsection{Evaluation of Introduction Cost}
In order to address the security shortcomings of embedded wallets, our introduced protocol's fundamental idea involves incorporating audit processes from an audit organization into the wallet system. The audit organization, which in this implementation is assumed to be a security audit company, is required to provide an audit contract built using smart contracts. Generally, developing and maintaining smart contracts requires high technical expertise, including scalable contract design and robust security measures. These requirements can be challenging for a typical organization to meet. In our approach, we utilize the Ethereum Name Service (ENS) to facilitate the audit contract. ENS, built on smart contract bases and managed by a Decentralized Autonomous Organization (DAO), is a distributed naming service. Using the TextRecord feature\cite{ENSIP5} in ENS, a mechanism for custom records, eliminates the need for developing audit contracts from scratch, thereby saving time and costs involved in implementation.

The costs necessary for the deployment of an audit contract using ENS are shown in Table 1. As all operations are performed via smart contracts, the costs include transaction gas fees and ENS fees. The initial setup cost, based on the ETH token price as of November 12, 2023, is below \$30.
\begin{table}[h]
\centering
\begin{tabular}{|l|r|r|}
\hline
\textbf{Action} & \textbf{Gas Used} & \textbf{USD} \\ \hline
ENS registration fee (1 year) & - & \$5.00 \\ \hline
Registration - Start timer & 44,206 & \$2.27 \\ \hline
Registration - Register name & 275,933 & \$14.16 \\ \hline
Update text records & 136,234 & \$7.01 \\ \hline
\textbf{Total Cost} & \textbf{-} & \textbf{\$28.44} \\ \hline
\end{tabular}
\caption{Costs for ENS and Related Operations (25 Gwei Gas Price, 1 ETH = \$2,057 on November 12, 2023)}
\label{tab:ens_costs_updated_nov12}
\end{table}

Regarding operational costs, while gas fees for transactions are not required for blockchain reference operations, maintaining blockchain nodes incurs costs. Depending on the setup, whether on-premises or using cloud services, costs may vary. A common cost-effective method is to use Ethereum JSON-RPC endpoints such as Infura\cite{infura_website}. According to Infura's pricing, up to 100,000 transactions per day are free, and 1 million transactions per day cost \$225 per month. This can be adjusted based on the scale, such as the number of users utilizing the wallet verifier.

Based on these calculations, it is demonstrated that audit organizations can feasibly construct audit contracts using ENS with realistic cost implications. Moreover, not only is the mechanism cost-effective, but since all registration operations on the audit contract are conducted via blockchain transactions, it is resistant to tampering of audit trails and offers high transparency, allowing anyone to access the audit trails.

\section{Related Work}
\textbf{Classification of Embedded Wallets.}
Erinle et al.\cite{erinle2023sok}, through an extensive literature survey, have established a comprehensive classification of cryptocurrency wallets, analyzing the unique characteristics and related security considerations for each wallet category. To the best of our knowledge, their work appears to provide the most comprehensive classification of wallets available to date. However, recent developments in wallet forms, namely embedded wallets, have not been adequately addressed in the literature. Although the term "embedded wallet" appears in recent work\cite{wang2022exploring} where the burgeoning concept of Web3 is explored from the perspective of blockchain architecture, this term describes wallets operating within a browser, akin to browser extensions such as Metamask, but does not delineate a distinct category. Conversely, in the industrial realm, embedded wallets are increasingly being recognized as a distinct form of cryptocurrency wallet offerings\cite{zengo2023personalvsembedded}. Therefore, in an academic context, our study appears to be the first to elucidate the structural features of embedded wallets and systematically categorize security concerns.

\textbf{Combatting Phishing and Scams.}
Research to prevent phishing scams related to blockchain projects has been increasing recently. In particular, scam account detection has been widely performed, with known methods including those based on traditional machine learning techniques specializing in feature extraction\cite{chen2020phishing}, and those using graph network embedding methods\cite{wu2020phishers, li2023siege}. However, in account-based detection, it is difficult to prevent phishing where the entire front-end behaves fraudulently. Therefore, approaches like those of Roy et al.\cite{roy2023unveiling}, which distinguish NFT phishing websites by features of the website and analyze phishing URLs to include them in browser protection tool blocklists, would be effective. 

However, these existing approaches, which mainly rely on identifying fraudulent activities through external observations, have limitations. Our research represents a first step forward in incorporating mechanisms into wallets that ensure authenticity and integrity, classic challenges in security. This approach is not intended to replace existing methods but to complement them.

\textbf{Auditing Blockchain Projects.}
Our approach leverages auditing practices used in Web3 and blockchain projects. As the idiom "Don’t trust, verify" is commonly used in the blockchain community, audits of blockchain projects are conducted extensively across various aspects. For example, Proof of Reserves in the audit of cryptocurrency exchange reserves is well known, and beyond this, various methods have been developed to provide auditability to individuals and regulatory authorities\cite{chatzigiannis2021sok}. Undergoing security audits at the launch of blockchain projects has become common for gaining user trust. In particular, academia has proposed many tools for smart contract security audits\cite{chaliasos2023smart}, and numerous blockchain-specialized security audit companies have been established in the industry, such as CertiK\cite{certik2023} and Hacken\cite{hacken2023}. In particular in the domain that our paper addresses, reports on the security audits of wallets have been published\cite{metamask_security2023}. On the other hand, the exploration of mechanisms for utilizing audit reports, as proposed by us, to verify audit subjects and alert users about potential authenticity and completeness deficiencies, remains an uncharted territory.

\section{Conclusion and Future Work}
This paper has clarified the security characteristics of recently emerging embedded wallets and discussed the potential to induce fraudulent transactions. As a solution, we proposed the VELLET protocol, a verifiable embedded wallet that incorporates the activities of audit organizations. Our proposal has been demonstrated to be implementable at a feasible cost through proof-of-concept.

Similar to embedded wallets, many blockchain projects prioritize user convenience and corporate profits, often at the expense of security. We believe that providing users with defenses against fraudulent Dapps is of utmost importance. In this study, we focused on securing wallet authenticity and integrity, utilizing by audits and  their trails as a key to user security. We consider this approach to be extendable not only to embedded wallets but also to various aspects of Dapps, for instance, to entire websites and smart contracts, and we have positioned it as a future challenge.

\bibliographystyle{IEEEtran}
\bibliography{main}

\begin{thebibliography}{10}
\providecommand{\url}[1]{#1}
\csname url@samestyle\endcsname
\providecommand{\newblock}{\relax}
\providecommand{\bibinfo}[2]{#2}
\providecommand{\BIBentrySTDinterwordspacing}{\spaceskip=0pt\relax}
\providecommand{\BIBentryALTinterwordstretchfactor}{4}
\providecommand{\BIBentryALTinterwordspacing}{\spaceskip=\fontdimen2\font plus
\BIBentryALTinterwordstretchfactor\fontdimen3\font minus \fontdimen4\font\relax}
\providecommand{\BIBforeignlanguage}[2]{{%
\expandafter\ifx\csname l@#1\endcsname\relax
\typeout{** WARNING: IEEEtran.bst: No hyphenation pattern has been}%
\typeout{** loaded for the language `#1'. Using the pattern for}%
\typeout{** the default language instead.}%
\else
\language=\csname l@#1\endcsname
\fi
#2}}
\providecommand{\BIBdecl}{\relax}
\BIBdecl

\bibitem{TripleA2023}
\BIBentryALTinterwordspacing
Triple-A, ``Cryptocurrency ownership data,'' accessed: 2023-11-23. [Online]. Available: \url{https://triple-a.io/crypto-ownership-data/}
\BIBentrySTDinterwordspacing

\bibitem{DappRadar2023}
\BIBentryALTinterwordspacing
DappRadar, ``{Dapps Industry Overview},'' accessed: 2023-12-03. [Online]. Available: \url{https://dappradar.com/industry-overview}
\BIBentrySTDinterwordspacing

\bibitem{metamask2023}
\BIBentryALTinterwordspacing
{MetaMask}, ``The crypto wallet for defi, web3 dapps and nfts,'' accessed: 2023-11-23. [Online]. Available: \url{https://metamask.io/}
\BIBentrySTDinterwordspacing

\bibitem{Froehlich2022}
M.~Fröhlich, F.~Waltenberger, L.~Trotter, F.~Alt, and A.~Schmidt, ``Blockchain and cryptocurrency in human computer interaction: A systematic literature review and research agenda,'' in \emph{Proceedings of the 2022 ACM Designing Interactive Systems Conference (DIS '22)}, 2022, pp. 155--177.

\bibitem{Voskobojnikov2021}
A.~Voskobojnikov, O.~Wiese, M.~M. Koushki, V.~Roth, and K.~Beznosov, ``The {U} in crypto stands for usable: An empirical study of user experience with mobile cryptocurrency wallets,'' in \emph{Proceedings of the 2021 CHI Conference on Human Factors in Computing Systems (CHI '21)}, 2021, pp. 1--14.

\bibitem{zengo2023personalvsembedded}
\BIBentryALTinterwordspacing
O.~Ohayon, ``Personal wallets vs. embedded wallets: Who wins in crypto?'' March 2023, accessed: 2023-11-23. [Online]. Available: \url{https://zengo.com/personal-wallets-vs-embedded-wallets-who-wins/}
\BIBentrySTDinterwordspacing

\bibitem{thirdweb2023embeddedwallet}
\BIBentryALTinterwordspacing
{Thirdweb}, ``Embedded wallets - overview,'' accessed: 2023-11-23. [Online]. Available: \url{https://portal.thirdweb.com/embedded-wallet}
\BIBentrySTDinterwordspacing

\bibitem{privy2023embeddedwallets}
\BIBentryALTinterwordspacing
{Privy}, ``Embedded wallets documentation,'' accessed: 2023-11-23. [Online]. Available: \url{https://docs.privy.io/guide/frontend/embedded/overview}
\BIBentrySTDinterwordspacing

\bibitem{dynamic2023embeddedwallets}
\BIBentryALTinterwordspacing
Dynamic, ``Overview of embedded wallets,'' 2023, accessed: 2023-11-23. [Online]. Available: \url{{https://docs.dynamic.xyz/embedded-wallets/overview}}
\BIBentrySTDinterwordspacing

\bibitem{ENSIP5}
\BIBentryALTinterwordspacing
R.~Moore, ``{ENSIP-5: Text Records},'' May 2017, {ENS Improvement Proposals, no. 5}. [Online]. Available: \url{https://docs.ens.domains/ens-improvement-proposals/ensip-5-text-records}
\BIBentrySTDinterwordspacing

\bibitem{ENSWebsite}
\BIBentryALTinterwordspacing
ENS, ``Ethereum name service: Decentralised naming for wallets, websites, \& more,'' accessed: 2023-11-23. [Online]. Available: \url{https://ens.domains}
\BIBentrySTDinterwordspacing

\bibitem{erinle2023sok}
Y.~Erinle, Y.~Kethepalli, Y.~Feng, and J.~Xu, ``Sok: Design, vulnerabilities, and security measures of cryptocurrency wallets,'' 2023, arXiv:2307.12874.

\bibitem{argent2023}
\BIBentryALTinterwordspacing
{Argent}, ``Argent – the best ethereum wallet for defi and nfts,'' 2023, accessed: 2023-11-23. [Online]. Available: \url{https://www.argent.xyz/}
\BIBentrySTDinterwordspacing

\bibitem{trustwallet2023}
\BIBentryALTinterwordspacing
{Trust Wallet}, ``Best crypto wallet for web3, nfts and defi,'' accessed: 2023-11-23. [Online]. Available: \url{https://trustwallet.com/}
\BIBentrySTDinterwordspacing

\bibitem{coinbase2023waas}
\BIBentryALTinterwordspacing
{Coinbase}, ``Waas - coinbase cloud,'' accessed: 2023-11-23. [Online]. Available: \url{https://www.coinbase.com/cloud/products/waas}
\BIBentrySTDinterwordspacing

\bibitem{circle2023programmablewallets}
\BIBentryALTinterwordspacing
{Circle}, ``Programmable wallets | wallet as a service,'' accessed: 2023-11-23. [Online]. Available: \url{https://www.circle.com/en/programmable-wallets}
\BIBentrySTDinterwordspacing

\bibitem{Chainalysis2023}
\BIBentryALTinterwordspacing
Chainalysis, ``{The Chainalysis 2023 Crypto Crime Report},'' 2023, accessed: 2023-11-23. [Online]. Available: \url{https://go.chainalysis.com/2023-crypto-crime-report.html}
\BIBentrySTDinterwordspacing

\bibitem{andryukhin2019phishing}
A.~A. Andryukhin, ``Phishing attacks and preventions in blockchain based projects,'' in \emph{2019 International Conference on Engineering Technologies and Computer Science (EnT)}, Moscow, Russia, 2019, pp. 15--19.

\bibitem{yang2023trail}
J.~Yang, J.~Liu, and J.~Wu, ``With trail to follow: Measurements of real-world non-fungible token phishing attacks on ethereum,'' 2023, arXiv:2307.01579.

\bibitem{trustwallet2023phishing}
\BIBentryALTinterwordspacing
{Trust Wallet Community}, ``How to spot a phishing attack \& protect your crypto,'' 2023, accessed: 2023-11-23. [Online]. Available: \url{https://community.trustwallet.com/t/how-to-spot-a-phishing-attack-protect-your-crypto/753663}
\BIBentrySTDinterwordspacing

\bibitem{metamask2023blockaid}
\BIBentryALTinterwordspacing
{MetaMask Support}, ``How to turn on blockaid security alerts,'' accessed: 2023-11-23. [Online]. Available: \url{https://support.metamask.io/hc/en-us/articles/19878220833947-How-to-turn-on-Blockaid-security-alerts}
\BIBentrySTDinterwordspacing

\bibitem{trustwallet2023securityscanner}
\BIBentryALTinterwordspacing
{Trust Wallet Community}, ``Introducing the trust wallet security scanner: Making crypto \& web3 safer for everyone,'' 2022, accessed: 2023-11-23. [Online]. Available: \url{https://community.trustwallet.com/t/introducing-the-trust-wallet-security-scanner-making-crypto-web3-safer-for-everyone/643056}
\BIBentrySTDinterwordspacing

\bibitem{rfc2535}
\BIBentryALTinterwordspacing
{D. E. Eastlake 3rd}, ``{Domain Name System Security Extensions},'' RFC 2535, Mar. 1999. [Online]. Available: \url{https://www.rfc-editor.org/info/rfc2535}
\BIBentrySTDinterwordspacing

\bibitem{cooper2018security}
\BIBentryALTinterwordspacing
D.~Cooper, A.~Regenscheid, M.~Souppaya, C.~Bean, M.~Boyle, D.~Cooley, and M.~Jenkins, ``Security considerations for code signing,'' \emph{NIST Cybersecurity White Paper}, 2018. [Online]. Available: \url{https://doi.org/10.6028/NIST.CSWP.01262018}
\BIBentrySTDinterwordspacing

\bibitem{microsoft_driver_signing}
Microsoft, ``Driver signing - windows drivers,'' \url{https://learn.microsoft.com/en-us/windows-hardware/drivers/install/} \url{driver-signing}, May 2023, accessed: 2023-11-23.

\bibitem{apple_code_signing}
Apple, ``About code signing,'' \url{https://developer.apple.com/library/archive/} \url{documentation/Security/Conceptual/CodeSigningGuide}, September 2016, accessed: 2023-11-23.

\bibitem{android_app_signing}
{Google for Developers}, ``Sign your app - android studio,'' \url{https://developer.android.com/studio/publish/app-signing}, 2023, accessed: 2023-11-23.

\bibitem{certik2023}
\BIBentryALTinterwordspacing
{CertiK}, ``Web3 security leaderboard,'' accessed: 2023-11-27. [Online]. Available: \url{https://www.certik.com/}
\BIBentrySTDinterwordspacing

\bibitem{hacken2023}
\BIBentryALTinterwordspacing
{Hacken}, ``Blockchain security services company - web3, crypto, defi,'' accessed: 2023-11-27. [Online]. Available: \url{https://hacken.io/}
\BIBentrySTDinterwordspacing

\bibitem{Weyl2022}
\BIBentryALTinterwordspacing
E.~G. Weyl, P.~Ohlhaver, and V.~Buterin, ``Decentralized society: Finding web3's soul,'' 2022, available at SSRN. [Online]. Available: \url{https://ssrn.com/abstract=4105763}
\BIBentrySTDinterwordspacing

\bibitem{EIP5192}
\BIBentryALTinterwordspacing
T.~Daubenschütz and Anders, ``{ERC-5192: Minimal Soulbound NFTs},'' July 2022, {Ethereum Improvement Proposals, no. 5192}. [Online]. Available: \url{https://eips.ethereum.org/EIPS/eip-5192}
\BIBentrySTDinterwordspacing

\bibitem{EIP721}
\BIBentryALTinterwordspacing
J.~E. William~Entriken, Dieter~Shirley and N.~Sachs, ``{ERC-721: Non-Fungible Token Standard},'' Jan 2018, {Ethereum Improvement Proposals, no. 721}. [Online]. Available: \url{https://eips.ethereum.org/EIPS/eip-721}
\BIBentrySTDinterwordspacing

\bibitem{infura_website}
\BIBentryALTinterwordspacing
Infura, ``Web3 development platform | ipfs api \& gateway | blockchain node service,'' 2023, accessed: 2023-11-23. [Online]. Available: \url{https://www.infura.io/}
\BIBentrySTDinterwordspacing

\bibitem{wang2022exploring}
Q.~Wang, R.~Li, Q.~Wang, S.~Chen, M.~Ryan, and T.~Hardjono, ``Exploring web3 from the view of blockchain,'' 2022, arXiv:2206.08821.

\bibitem{chen2020phishing}
W.~Chen, X.~Guo, Z.~Chen, Z.~Zheng, and Y.~Lu, ``Phishing scam detection on ethereum: Towards financial security for blockchain ecosystem.'' in \emph{IJCAI}, vol.~7, 2020, pp. 4456--4462.

\bibitem{wu2020phishers}
J.~Wu, Q.~Yuan, D.~Lin, W.~You, W.~Chen, C.~Chen, and Z.~Zheng, ``Who are the phishers? phishing scam detection on ethereum via network embedding,'' \emph{IEEE Transactions on Systems, Man, and Cybernetics: Systems}, vol.~52, no.~2, pp. 1156--1166, 2020.

\bibitem{li2023siege}
S.~Li, R.~Wang, H.~Wu, S.~Zhong, and F.~Xu, ``Siege: Self-supervised incremental deep graph learning for ethereum phishing scam detection,'' in \emph{Proceedings of the 31st ACM International Conference on Multimedia}, 2023, pp. 8881--8890.

\bibitem{roy2023unveiling}
S.~S. Roy, D.~Das, P.~Bose, C.~Kruegel, G.~Vigna, and S.~Nilizadeh, ``Unveiling the risks of nft promotion scams,'' 2023, arXiv:2301.09806.

\bibitem{chatzigiannis2021sok}
P.~Chatzigiannis, F.~Baldimtsi, and K.~Chalkias, ``Sok: Auditability and accountability in distributed payment systems,'' in \emph{International Conference on Applied Cryptography and Network Security}.\hskip 1em plus 0.5em minus 0.4em\relax Springer, 2021, pp. 311--337.

\bibitem{chaliasos2023smart}
S.~Chaliasos, M.~A. Charalambous, L.~Zhou, R.~Galanopoulou, A.~Gervais, D.~Mitropoulos, and B.~Livshits, ``Smart contract and defi security: Insights from tool evaluations and practitioner surveys,'' 2023, arXiv:2304.02981.

\bibitem{metamask_security2023}
\BIBentryALTinterwordspacing
{MetaMask}, ``Security bug bounties,'' accessed: 2023-11-27. [Online]. Available: \url{https://metamask.io/security/}
\BIBentrySTDinterwordspacing

\end{thebibliography}

\end{document}